\documentclass[sigconf, nonacm]{acmart}

\newcommand\vldbyear{2025}
\newcommand\vldbworkshop{Tabular Data Analysis (TaDA)}
\newcommand\vldbauthors{\authors}
\newcommand\vldbtitle{\shorttitle} 
\newcommand\vldbavailabilityurl{https://github.com/IBM/ContextAwareJoin}
\newcommand\vldbpagestyle{plain} 
\usepackage{xcolor}

\usepackage{subcaption}
\usepackage{graphicx}
\usepackage{algorithm}
\usepackage{algpseudocode}
\usepackage{multirow}

\usepackage[capitalise,nameinlink]{cleveref}

\usepackage[framemethod=default]{mdframed}

\usepackage{colortbl}
\definecolor{bgcolor}{RGB}{240,240,240}
\newmdenv[hidealllines=true,backgroundcolor=bgcolor,innerleftmargin=.5em,innerrightmargin=.5em,innertopmargin=.2em,innerbottommargin=.8em,skipbelow=\baselineskip,skipabove=\baselineskip]{callout}

\newcommand{\industry}[1]{{{#1}}}

\newtheorem{example}{Example}
\newtheorem{definition}{Definition}

\newcommand{\introparagraph}[1]{\noindent\textbf{#1.}} %
\usepackage{booktabs}
\usepackage{ifthen}
\newcommand\ourmethod{TOPJoin}

\begin{document}
\title[TOPJoin: A Context-Aware Multi-Criteria Approach for Joinable Column Search]{TOPJoin: A Context-Aware Multi-Criteria Approach \\ for Joinable Column Search}

\author{Harsha Kokel}
\orcid{0000-0002-7548-3719}
\affiliation{%
  \institution{IBM Research}
}
\email{harsha.kokel@ibm.com}
\author{Aamod Khatiwada}
\authornote{Work done while at IBM}
\orcid{0000-0001-5720-1207}
\affiliation{%
  \institution{Northeastern University}
}
\email{khatiwada.a@northeastern.edu}

\author{Tejaswini Pedapati}
\orcid{0000-0002-5260-0951}
\affiliation{%
  \institution{IBM Research}
}
\email{tejaswinip@us.ibm.com}

\author{Haritha Ananthakrishnan}
\orcid{0009-0007-9906-2998}
\affiliation{%
  \institution{IBM Research}
}
\email{hananthakris@ibm.com}

\author{Oktie Hassanzadeh}
\orcid{0000-0001-5307-9857}
\affiliation{%
  \institution{IBM Research}
}
\email{hassanzadeh@us.ibm.com}

\author{Horst Samulowitz}
\orcid{0000-0002-6780-3217}
\affiliation{%
  \institution{IBM Research}
}
\email{samulowitz@us.ibm.com}

\author{Kavitha Srinivas}
\orcid{0000-0003-4610-967X}
\affiliation{%
  \institution{IBM Research}
}
\email{kavitha.srinivas@ibm.com}

\begin{abstract}
    One of the major challenges in enterprise data analysis is the task of finding joinable tables that are conceptually related and provide meaningful insights. Traditionally, joinable tables have been discovered through a search for similar columns, where two columns are considered similar syntactically if there is a set overlap or they are considered similar semantically if either the column embeddings or value embeddings are closer in the embedding space.
    However, for enterprise data lakes, column similarity is not sufficient to identify joinable columns and tables. The context of the query column is important. Hence, in this work, we first define \emph{context-aware column joinability}. Then we propose a multi-criteria approach, called TOPJoin, for joinable column search. We evaluate TOPJoin against 
    existing join search baselines over one academic and one real-world join search benchmark. Through experiments, we find that TOPJoin performs better on both benchmarks than the baselines. 
\end{abstract}

\maketitle

\pagestyle{\vldbpagestyle}
\begingroup\small\noindent\raggedright\textbf{VLDB Workshop Reference Format:}\\
\vldbauthors. \vldbtitle. VLDB \vldbyear\ Workshop: \vldbworkshop.\\ %
\endgroup
\begingroup
\renewcommand\thefootnote{}\footnote{\noindent
This work is licensed under the Creative Commons BY-NC-ND 4.0 International License. Visit \url{https://creativecommons.org/licenses/by-nc-nd/4.0/} to view a copy of this license. For any use beyond those covered by this license, obtain permission by emailing \href{mailto:info@vldb.org}{info@vldb.org}. Copyright is held by the owner/author(s). Publication rights licensed to the VLDB Endowment. \\
\raggedright Proceedings of the VLDB Endowment. %
ISSN 2150-8097. \\
}\addtocounter{footnote}{-1}\endgroup

\ifdefempty{\vldbavailabilityurl}{}{
\vspace{.3cm}
\begingroup\small\noindent\raggedright\textbf{VLDB Workshop Artifact Availability:}\\
The source code, data, and/or other artifacts have been made available at \url{\vldbavailabilityurl}.
\endgroup
}

\section{Introduction}

With escalating enterprise dataset sizes, there's a pressing need to automate data analysis. Data Discovery is a significant hurdle in automating data analysis.
\industry{
In simpler terms, data discovery is the process of finding relevant data sources to answer specific business questions. 
In the context of building a conversational interface for Semantic Automation Layer~\cite{WeideleRBVBCMMA23,Mihindukulasooriya23,WeideleMVRSFAMA24}, this paper focuses on a specific aspect of data discovery: \emph{joinable table search} i.e., given a set of tables and a query table with a designated join column, identify the tables and additional columns (features) that can 
expand the query table horizontally and generate valuable insights~\cite{ALITE_khatiwadaSGM22}. 
}

Conventionally, Joinable tables have been identified by comparing their columns' similarity. Two columns are deemed syntactically similar if there values overlap~
~\cite{JOSIE_ZhuDNM19, LSHENSEMBLE_ZhuNPM16}
or semantically similar if their column embeddings or value embeddings are closer in the embedding space~
\cite{PEXESO_DongT0O21, DeepJoin_Dong0NEO23, WarpGate_CongGFJD23}.
However, focusing solely on column similarity is insufficient for enterprise data lakes as they contain unrelated data from diverse sources. 
\industry{This is a critical consideration for industry applications, as false positives resulting from joins on columns with shared values but lacking contextual relevance can overwhelm users and hinder the discovery of valuable insights. So, it is imperative to minimize false positive joins and prioritize joins that are both statistically significant and contextually relevant.}
Consider the following example from the Open Data,  which is widely used for data discovery~\cite{datalake_tutorial_NargesianZMPA19}. 

\begin{table*}[t]
\centering
\vspace{-2em}
 \caption{Open Data Tables.}
    \includegraphics[width=\textwidth]{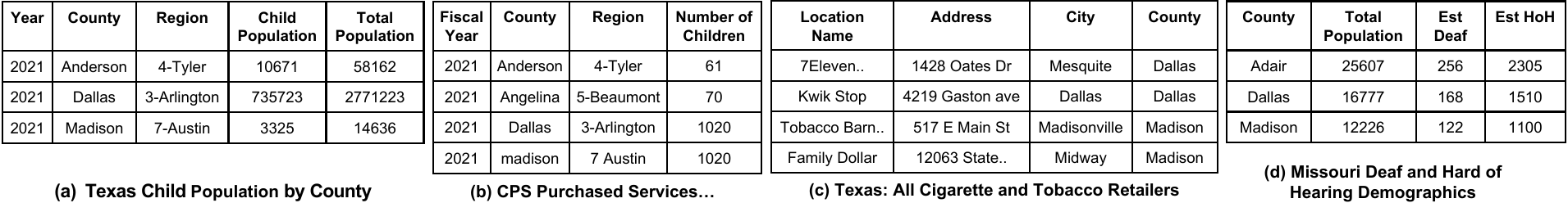}
    \vspace{-2em}
\label{fig:topjoin_main}
\end{table*}

\begin{callout}

\begin{example}\label{ex:context_aware_join}

\small
\cref{fig:topjoin_main}(a,b)
are about the child population in Texas,~\cref{fig:topjoin_main}(c) is about tobacco retailers in Texas, while ~\cref{fig:topjoin_main}(d) is about Missouri. Assume Table \texttt{a.County} (Table I(a) and Column `County') is the query, typical joinability criteria, based on value and semantic overlaps, identifies 
Tables \texttt{b.County, c.County} and \texttt{d.County} to be joinable with the query column.
However, from the user's perspective, joining~Tables (a) and (b) could make more sense as it augments the query table to provide insights
such as what proportion of the child population received assistance. Joining Tables (a) and (c) maybe less preferable because the join will increase the size of the original query table (each row would get repeated for each retailer in that county). Joining Tables (a) and (d) does not yield additional insights; in fact, the join is not sensible because the places are not the same. So, labeling Table (d) as joinable does not aid in data discovery. 
Furthermore, the representation of the joinable values could be different
(e.g., \verb+Madison+ in Table (a) vs \verb+madison+ in Table (b), or NY vs New York). 
Ideally, any computation of joinability should consider %
such fuzzy joins as well.

\end{example}
\end{callout}

\cref{ex:context_aware_join} highlights the following \emph{key factors} for join search: (1) It is essential to consider the context of the query to ensure meaningful results. (2) Smaller join sizes are generally preferred for effective data analytics.
(3) It is important to consider fuzzy joins, which allow for some degree of mismatch or require transformation before joining.
Motivated by these key factors, we first propose a new definition of \emph{context-aware joinable columns} and then present a multi-criteria approach, called \emph{TOPJoin} 
to tackle the challenges of joinability and ensure that the joined tables are meaningful.  
TOPJoin consolidates the rankings of various preference criteria, each assessing the column's joinability based on distinct property. 
Preference criteria includes degrees of column value overlap and unique values in the column, the closeness of column metadata's embeddings, the closeness of column values' embeddings, and potential output sizes of the table if the join was performed on the column.
Particularly, column value embeddings
facilitate the fuzzy joins discussed in the third key factor. 
However, using embeddings can be misleading as well. Strings that are not the same entity may still be similar in an embedding space (e.g., Boston and Brockton).  
We therefore need to weight each preference criterion carefully; exact matches are clearly better than embedding matches. 
We use a standard multi-criteria decision-making technique to perform this weighted preference ranking~\cite{TOPSIS_HwangY81}.

To show the effectiveness, we tested \emph{TOPJoin} on 
one enterprise and one academic benchmark, and compared it with existing approaches, including the conventional approach of 
ranking joinable columns based on their value overlap~\cite{JOSIE_ZhuDNM19}. In both cases, \emph{TOPJoin} showed the best performance.  

\section{Problem Definition}

Two tables can be joined 
if each of them contains
a column that can be joined.
Consequently,
the problem of discovering join tables is formulated as a \emph{joinable column search problem} defined as follows:
\begin{definition}\label{def:join_search}
    \textbf{Joinable Column Search Problem:} Given a collection of columns $\mathcal{C}$, a query column $q_Q$ from a table $Q$, and a constant $k$, find top-$k$ columns from the collection that are \emph{``joinable"} to the query column $q_Q$.
\end{definition}
\noindent
While different notions of \emph{``joinable"} columns have been explored 
in the literature
~\cite{datalake_tutorial_NargesianZMPA19}, %
popular definitions are based on set similarity.
These set-similarity based approaches, measures set intersection size~\cite{JOSIE_ZhuDNM19}, jaccard similarity~\cite{Mannheim_LehmbergRRMPB15}, or set containment~\cite{LSHENSEMBLE_ZhuNPM16} between a pair of columns. 
Approaches that use such syntactic join measures are often referred as \emph{equality join} or \emph{equi-join} approaches.
Some recent work explores 
the notion of 
\emph{semantically-joinable}. Specifically, \citet{PEXESO_DongT0O21} defined semantic joinability as a measure of similarity between the embeddings of values in the column (they call it vector matching). \citet{WarpGate_CongGFJD23} studied the semantic column joinability defined as similarity between the column embeddings. 
\citet{Valentine_KoutrasSIPBFLBK21} defined column pairs as semantically joinable if the columns share semantically equivalent instances.
Notably, none of these approaches consider table context in the joinable column definition. In our work, as shown in Example~\ref{ex:context_aware_join}, we are motivated to find joinable columns based on table context. So we define \emph{context-aware joinable column} as follows.

\begin{definition}
    \textbf{Context-aware Joinable Columns:} Given a query column $q_Q$ and a target column $t_T$, the following two conditions must be met for the columns to be context-aware joinable:
        (i) The tables $Q$ and $T$ must possess a semantic relationship or belong to related domains, indicating that the data in these tables are connected or relevant to one another. 
        (ii) The columns $q_Q$ and $t_T$ must be semantically joinable, meaning that they can be combined to produce meaningful results that satisfy the query's objectives.
\end{definition}

\industry{This notion of context-aware Joinable Columns is especially important in enterprise applications, where the data lakes may contain tables from various organizations and the user queries are intended for discovering insights.}
For evaluating techniques based on the definition, we generate ground truth by using diverse annotations that capture various semantic relationships between tables which we discuss in \cref{sec:datasets}. In \cref{sec:topjoin} we present our proposed multi-criteria approach.

\section{\ourmethod{}}\label{sec:topjoin}

One common method used in multi-criteria decision-making is called \emph{Technique for Order of Preference by Similarity to Ideal Solution} (TOPSIS)~\cite{TOPSIS_HwangY81}.
Given a set of candidates and different preferences, TOPSIS ranks the candidates based on the distances from an ideal solution.
Specifically, the TOPSIS procedure begins by establishing a normalized decision matrix, where each row represents a candidate and each column represents a preference criterion. If the criteria carry unequal weights, the matrix undergoes reweighting to reflect this. Following this, 
the ideal solution is defined as an instance with the best possible value for each criterion. 
Subsequently, the distance of each candidate from the ideal solution is calculated, leading to the candidates being ranked accordingly.

To improve the search for joinable columns across multiple preference criteria, we employ the TOPSIS method to retrieve the ranked results for column search. Our method, \emph{Technique for Order of Preference for Join Search} (\ourmethod{}) ,
works in two steps:
(1) It identifies potential join candidates using three different strategies---discussed in \cref{sec:candidates}. 
(2) It combines the identified candidates and ranks them using TOPSIS by consolidating different preference criteria discussed in \cref{sec:rankers}.

\subsection{Candidate Identification Strategies}\label{sec:candidates}

To identify candidate columns for context-aware joinable column search, we employ three distinct strategies. Our first strategy 
utilizes
\emph{syntactic joinability measures}, which have been widely adopted by various approaches (cf \cite{JOSIE_ZhuDNM19,LSHENSEMBLE_ZhuNPM16,LSHForest_BawaCG05}). In our experiments, to determine the columns that have maximum overlap, we implement
JOSIE~\cite{JOSIE_ZhuDNM19}. Essentially, we build a posting list for each normalized cell value which points to a list of columns that contains that value. 
Then we store the posting list in the form of an inverted index.
During search time, we utilize the inverted index
to pinpoint the columns exhibiting the highest degree of syntactic joinability. Significant storage is required to store indexes for large data lakes that may exceed product stipulations.
Therefore, we use approx. indexing techniques.

In our second strategy, we utilize a \emph{metadata semantics} technique to identify potential table-column join candidates that are semantically related to the query table and column. Our method entails crafting a metadata sentence for each column by combining various metadata elements such as the table name, table description, column details, dataset source, the query column's name and description, etc. Subsequently, employing a sentence transformer~\cite{sbert}, we produce sentence embeddings for each column and establish an index. When conducting searches, we employ K-nearest neighbor search within the embedding space to identify candidate columns sharing similar contextual attributes with the query's metadata.

Lastly, in our third strategy, \emph{value semantics}, we transform each data lake column into text by concatenating its values, and embed it using a sentence transformer. We similarly embed the query column and search for data lake columns (join candidates) having embeddings
that are closest to the query column.

\vspace{-1em}
\subsection{Preference Criteria}\label{sec:rankers}

\introparagraph{\textbf{Unique Values}} 
Generally, in databases, 
we 
join a primary key column with a foreign key column.
Importantly, a column is assigned a primary key if it has all unique values.
A column with a higher percentage of unique values has a higher probability of being a primary key column.
Thus, we rank the columns based on unique values, scored by the ratio between the number of distinct values in a column to the total number of rows in a column.

\introparagraph{\textbf{Intersection Size}}\label{sec:pc_intersection_size} 
  An ideal data lake column for joining
  must have overlapping values with the query column. This can be computed using exact matches from posting lists.  However, as it is computationally prohibitive to compute the overlap between columns in large enterprise data lakes, we represent each column by its minhash values~\cite{LSHForest_BawaCG05}. The intersection size is now determined by the hamming distance between these minhash representations.

\introparagraph{\textbf{Join Size} and \textbf{Reverse Join Size}}
While joining tables in
~\cref{fig:topjoin_main} (a) and (b),
the overall cardinality of the joined table is the same as the original tables (a) and (b), and it also provides additional insights. 
However, if we combine the tables in
~\cref{fig:topjoin_main} (b) and (c), 
the cardinality of the resulting table is much larger than the original sizes.
Furthermore, the columns \texttt{Region}, \texttt{Child Population} and \texttt{Total Population} in the resulting join table would be duplicated across rows. 
In an ideal join scenario, we want the cardinality of the joined table to be reasonable and also avoid rows with duplicate values in some columns. 
Given a query and a candidate table, \textbf{Join size} is the size of the combined table when left join is performed and \textbf{reverse join size} is the size of the combined table when right join is performed. 
We estimate the cardinality of the joined table based on a prior work~\cite{JoinSize_Swami94}.

\introparagraph{\textbf{Value Semantics}} 
\label{sec:pc_value_semantics}
  To understand if the query column and the candidate column are joinable, the most common strategy is to calculate their intersection size.
If there is a high overlap among them, then the columns are deemed joinable. 
However, just considering the raw values or their minhash representation fails to capture the overall concepts of both the columns and see if the columns are semantically similar or related.
For instance, consider two columns, one has values \textbf{New York, Los Angeles, San Francisco, Washington D.C.} and another has values \textbf{NYC, LA, SF, DC} etc. 
Their intersection 
yield a null set indicating that they should not be joined. However, these values indicate the same real-world entities, one encoding them in full and the next abbreviating them.
Thus, to take the semantics or concepts of each column into consideration, we first compute the most frequent values of that column.
We transform the list of most frequent values to text (a sentence) and obtain embedding for that text.
The similarity of the concepts in the two columns is then quantified as the cosine similarity of the frequent value embeddings of the two columns.
We only use the value semantics criterion for string columns as integers/floats are less likely to provide semantic information.

\introparagraph{\textbf{Disjoint Value Semantics}}
  Value semantics considers the concepts in the entire column. If the query column and the candidate column share a lot of values, then their concepts would naturally be similar. 
  It is crucial to confirm that values not shared by the query and the candidate still have similar meanings.
To evaluate this criterion, we consider two sets: (1) the values that are present in the query column but not in the candidate column; (2) the values that are present in the candidate column and not in the query column. 
We then generate the embeddings for the values for both these sets and compute their cosine similarity.
If the similarity is high, it indicates that the columns are indeed related (e.g. one column could be the abbreviation of the other). 
Like value semantics, we use the disjoint value semantics criterion 
only 
for string columns.

\introparagraph{\textbf{Metadata Semantics}}
  In addition to utilizing the values and its concepts for a given column, we also leverage the additional metadata provided. 
This could include the table description, the tags describing the table, the column description, 
column names in the table,
etc. 
We obtain the embeddings of this metadata and compute the similarity between the embeddings of both columns.

Along with column values, TOPJoin relies on their other representations such as minhash, metadata embeddings, and frequent value embeddings.
To improve search time efficiency,
we compute these representations of the data lake columns
during a pre-processing phase and index them. 
\industry{In most applications, this indexing happens when a new table is added to the data lake.}
In the search phase, we look up the details for the query and use the index to obtain the candidates.
Moreover, most common database indexes also have built-in functions to compute hamming distance, cosine similarity, etc. As some of our preferences rely on computing the hamming distance and cosine similarity between two vectors, we leverage these native functions to accelerate the search. We provide further indexing details in the experiments~(\cref{sec:experiments}).

\section{Experiments}\label{sec:experiments}

For empirical evaluation, we 
randomly select $1$ million values for a column if it has more values.
While indexing, \ourmethod{} builds three indexes for three candidate identification strategies discussed in~\cref{sec:candidates}. 
For syntactic joinability, we utilize an inverted index.
For the other two strategies, we use 
a bert-based Sentence Transformer 
to generate the embeddings and leverage standard nearest neighbors for indexing (Sklearn~\cite{scikit-learn} NearestNeighbors indexer with default parameters). During the search process, we retrieve the top-$100$ candidates from each of these indexes and compute preferences on the merged set. 
TOPSIS permits us to assign varying weights to each preference criterion. 
\industry{However, for industry applications, it is important to have a consistent performance across clients.} Hence, 
in all our assessments, we allocated a weight of 0.2 to every criterion, except for the intersection size, which was assigned a weight of 0.5 based on experimental results. 

\begin{table}[t]
\caption{Comparison of \ourmethod{} with LSH Ensemble and DeepJoin; at K=10. Best results are indicated in \textbf{bold} and second best are \underline{underlined}.}
\label{tab:topjoin_short_results}
\vspace{-1em}
\centering
\renewcommand{\arraystretch}{0.4}
\resizebox{1\linewidth}{!}{
\begin{tabular}{c||c|c|c||c|c|c}
\toprule
\multirow{2}{*}{\textbf{Method}} & \multicolumn{3}{c||}{\textbf{CIO}}  & \multicolumn{3}{c}{\textbf{Open Data}}  \\   
    & \textbf{MRR}  & \textbf{MAP} &  \textbf{Recall} & \textbf{MRR}  & \textbf{MAP} &  \textbf{Recall}   \\ 

\midrule
{LSH Ensemble} &0.29 & 0.25 & 0.41   &0.44 &0.37 &0.46  \\
\midrule
{DeepJoin} & 0.20 & 0.16 & 0.29 & 0.44 & 0.36 & 0.47 \\
\midrule
{\ourmethod{}} & \textbf{0.39} & \textbf{0.34} & \textbf{0.68} & \textbf{0.51} & \textbf{0.49} & \textbf{0.65} \\
\midrule
{\ourmethod{}-Minhash} & \underline{0.37} & \underline{0.33} & \underline{0.60} & \underline{0.60} & \underline{0.48} & \underline{0.62} \\ 
\bottomrule
\end{tabular}
} 
\vspace{-1em}
\end{table}

\vspace{-0.2em}
\subsection{Datasets}\label{sec:datasets}

To demonstrate that table context is important while identifying joins and that our approach is effective, 
we include one real-world datalake and one academic benchmark.

\introparagraph{CIO real-world datalake} 
The CIO datalake is a replica of a \textbf{real-world} enterprise Operational Data Store (ODS)~\cite{inmon1995building}.
This datalake houses a set of database tables that contain transactional data that are derived from a number of DB2 databases and other NoSQL sources.
The dataset includes $59$ tables, and the ground truth for joinable columns was derived by analyzing a collection SQL query logs maintained in the ODS. The datalake features abbreviated column names and a mix of string, integer, float, and date columns. The number of rows in the tables varies from $0$ to $90$ million, with an average 
of $5,163,737$.

\introparagraph{OpenData}
The OpenData benchmark is a collection of tables extracted from open government data repositories, which have been frequently used in academic literature for set-similarity-based join search studies~\cite{JOSIE_ZhuDNM19,LSHENSEMBLE_ZhuNPM16}. In prior works, an exact method was utilized to determine set overlap and leverage it as a benchmark for joinability evaluation. This can be a problem as illustrated in \cref{ex:context_aware_join}. 
In our work, we gather \emph{human annotations for this dataset to enable the creation of context-aware joins that reflect real-world user behavior}.
We identified $471$ column pairs that exhibited high containment scores and solicited join labels from fifteen human annotators.
During the annotation process, human annotators had access to table snippets, metadata such as table descriptions, organization IDs, and tags, as well as the potential outcome of the joined table.

We received $6$ to $15$ annotations per sample and regarded columns as joinable if at least $10\%$ of the annotations were positive. 
Subsequently, out of $471$ pairs only only $42$ were identified as context-aware joinable column pairs.

\vspace{-0.2em}
\subsection{Results}\label{sec:results}

We compare \ourmethod{} with an equality join approach---LSH Ensemble---and a semantic join approach---DeepJoin. 

\textbf{LSH Ensemble}~\cite{LSHENSEMBLE_ZhuNPM16} identifies the top-K joinable columns in a data lake by using set containment as a criterion for joinability. It approximates the containment score between column pairs through the use of the Locally Sensitive Hashing (LSH) Ensemble Index. Our experiments use the publicly available implementation of LSH Ensemble with default parameters. 

\textbf{DeepJoin}~\cite{DeepJoin_Dong0NEO23} is an embedding-based approach utilizes approximate nearest neighbor search for retrieval. They transform columns into text and then into embeddings using various methods, including Pretrained Language Model (PLM) such as FastText~\cite{FastText_MikolovGBPJ18}, BERT~\cite{BERT_DevlinCLT19}, and MPNet~\cite{MPNet_Song0QLL20} as well as fine-tuned DistilBERT and MPNet. The resulting embeddings are then indexed using HNSW~\cite{HNSW_MalkovY20} for efficient search. Due to the unavailability of finetuned model and the training data, we have replicated DeepJoin model based on the publicly available implementation%
by training it on OpenData.

The syntactic join candidate identification strategy described in~\cref{sec:candidates} maintains an inverted index for cell values. \industry{The size of this inverted index can grow quickly for large data lakes. In our experiments, we limit the size by sampling 10K rows from the tables. 
We, further, examine the utility of an approximate approach (similar to what we used for the intersection size preference criterion (\cref{sec:pc_intersection_size})).}
As minhashes can efficiently estimate Jaccard similarity, we compute minhashes for each column (with $100$ permutations) and create an index using the hamming distance. At search time, we query the index for $k$ closest columns and use that as syntactic join candidates. We continue to utilize samples of original column values to compute the Minhashes. However, we maintain only the Minhash index instead of the inverted index. This choice leads to significant reductions in memory requirements. For instance, \industry{the inverted index for the CIO column, which had a size of $2.6$GB, is replaced by a minhash index that is only $508$KB in size. This reduction in memory usage is particularly beneficial for handling large datasets, particularly for enterprise applications. We refer to this approximate version as \textbf{\ourmethod-Minhash}.}

We present the Mean Reciprocal Ranks (MRR), Mean Average Precision (MAP), and Recall at K=10 in \cref{tab:topjoin_short_results}. 
Across both the datasets, we see that \ourmethod{} performs the best. While using the approximation, in \ourmethod-Minhash, does impact the MAP and Recall values, the performance is still satisfactory given the significant reduction in the storage requirements.

\section{Conclusion}

We introduced a novel problem of \emph{Context-aware} Joinable Column search \industry{that transcends traditional notions of semantic and syntactic joins. This notion of joinable column addresses the complex requirements of enterprise applications that targets more sophisticated data and insight discovery capabilities like Semantic Automation Layer~\cite{WeideleRBVBCMMA23,Mihindukulasooriya23,WeideleMVRSFAMA24}.}  
We presented \emph{\ourmethod{}}, a novel system that considers table-context and multiple preference-criteria in a principled way to find joinable tables. 

We conducted a study by deploying our system on an industrial-scale enterprise database and open data lake with manually annotated ground truth, to demonstrate its practical effectiveness in real-world settings.
Our results show that our system provides better results than previous work that consider only a single criteria of column value similarity. Current techniques used in \ourmethod{} for identifying semantically similar candidates is not finetuned for tabular data. There is enough evidence that suggest that models finetuned for tabular data are better for data-discovery task than the pretrained models. So adapting the embedding-based approached finetuned on tabular data as one of the candidate strategy in the \ourmethod{} could be a good future work.

\bibliographystyle{ACM-Reference-Format}
\bibliography{biblio}

\end{document}